\documentclass{appolb}
\usepackage{graphicx}
\usepackage{bm}
\usepackage{amssymb,amsmath,latexsym}
\usepackage{color}
\definecolor{darkblue}{RGB}{0,0,196}
\definecolor{darkred}{RGB}{196,0,0}
\usepackage{array}
\usepackage[sort&compress,numbers,comma]{natbib}

\begin{document}
\title{Bottomonium suppression in the quark-gluon plasma -- From effective field theories to non-unitary quantum evolution%
\thanks{Presented at Excited QCD 2022, Giardini Naxos, Sicily, October 2022}%
}
\author{Michael Strickland
\address{Department of Physics, Kent State University, Kent, OH 44242, USA}}
\maketitle
\begin{abstract}
In this proceedings contribution I review recent work which computes the suppression of bottomonium production in heavy-ion collisions using open quantum systems methods applied within the potential non-relativistic quantum chromodynamics (pNRQCD) effective field theory. I discuss how the computation of bottomonium suppression can be reduced to solving a Gorini-Kossakowski-Sudarshan-Lindblad (GKSL) quantum master equation for the evolution of the $b\bar{b}$ reduced density matrix.  The open quantum systems approach used allows one to take into account the non-equilibrium dynamics and decoherence of bottomonium in the quark-gluon plasma.  Finally, I present comparisons of phenomenological predictions obtained using a recently obtained next-to-leading-order GKSL equation with ALICE, ATLAS, and CMS experimental data for bottomonium suppression and elliptic flow.
\end{abstract}
  
\section{Introduction}

Heavy-ion collisions have been used to produce and study the properties of the quark-gluon plasma (QGP), a state of matter thought to have existed in the early universe and being created terrestrially in relativistic heavy-ion collisions. The suppression of bottomonium production in such collisions is considered strong evidence for the creation of a deconfined QGP~\cite{STAR:2013kwk,PHENIX:2014tbe,STAR:2016pof,CMS:2017ycw,Sirunyan:2018nsz,ALICE:2019pox,Acharya:2020kls,Lee:2021vlb,CMS:2020efs,CMS:2022rna}.
In the past, it was proposed that this suppression was due to the Debye screening of chromoelectric fields in the QGP, which modified the potential between heavy quarks and resulted in a reduction of heavy-quarkonium production~\cite{Matsui:1986dk,Karsch:1987pv}. However, more recent studies have shown that, in addition to the real part of the potential being modified by Debye screening, there is also an imaginary contribution to the potential caused by processes such as Landau damping and singlet-to-octet transitions~\cite{Laine:2006ns,Brambilla:2008cx,Beraudo:2007ky,Escobedo:2008sy,Dumitru:2009fy,Brambilla:2010vq,Brambilla:2011sg,Brambilla:2013dpa}. These processes result in large in-medium widths for heavy-quarkonium bound states.

In the past decade, there has been significant progress in the use of open quantum systems (OQS) methods to study heavy-quarkonium suppression in the QGP~\cite{Akamatsu:2011se,Akamatsu:2014qsa,Blaizot:2015hya,Brambilla:2016wgg,Blaizot:2017ypk,Brambilla:2017zei,Blaizot:2018oev,Yao:2018nmy,Miura:2019ssi,Brambilla:2019tpt,Sharma:2019xum,Akamatsu:2020ypb,Yao:2020xzw,Yao:2020eqy,Blaizot:2021xqa,Yao:2021lus,Katz:2015qja}. In particular, recent works have applied OQS methods within the framework of the potential non-relativistic QCD (pNRQCD) effective field theory~\cite{Pineda:1997bj,Brambilla:1999xf,Brambilla:2004jw,Brambilla:2008cx,Escobedo:2008sy,Brambilla:2010vq,Brambilla:2011sg,Brambilla:2013dpa}. The pNRQCD EFT is applicable to systems with a large separation between energy scales. This naturally occurs when the velocity of the heavy quark relative to the center of mass is small ($v \ll 1$). In Refs.~\cite{Brambilla:2016wgg,Brambilla:2017zei,Brambilla:2019tpt}, the authors considered the scale hierarchy relevant for small bound states in a high-temperature QGP, $1/r \sim Mv \gg m_D \sim \pi T \gg E$, where $r$ is the typical size of the state, $M$ is the heavy quark mass, $m_D$ is the Debye mass, $T$ is the temperature, and $E$ is the binding energy.  

With this scale ordering the environment's relaxation timescale is much shorter than both the system's internal timescales and the system's own relaxation timescale.  This makes the quantum evolution Markovian.  In Ref.~\cite{Brambilla:2019tpt} a Markovian Gorini-Kossakowski-Sudarshan-Lindblad (GKSL) equation~\cite{Gorini:1975nb,Lindblad:1975ef} was derived for the heavy-quarkonium reduced density matrix, which was implemented in the open-source~\texttt{QTraj} code of Ref.~\cite{Omar:2021kra} to make predictions for heavy-ion collision bottomonium observables~\cite{Brambilla:2020qwo,Brambilla:2021wkt,Brambilla:2022ynh,Alalawi:2022gul}. This was done by coupling the GKSL solver to 3+1D viscous hydrodynamics code using smooth Glauber initial conditions \cite{Alqahtani:2017mhy,Alqahtani:2020paa,Alalawi:2021jwn} and, most recently, fluctuating hydrodynamical backgrounds~\cite{Alalawi:2022gul}. The formalism used in the most recent works \cite{Brambilla:2022ynh,Alalawi:2022gul} is accurate to next-to-leading order (NLO) in the binding energy over temperature, which allows it to be used at lower temperatures than the original leading-order formalism. 

\section{Results}

\begin{figure}[t]
\begin{center}
\includegraphics[width=0.48\linewidth]{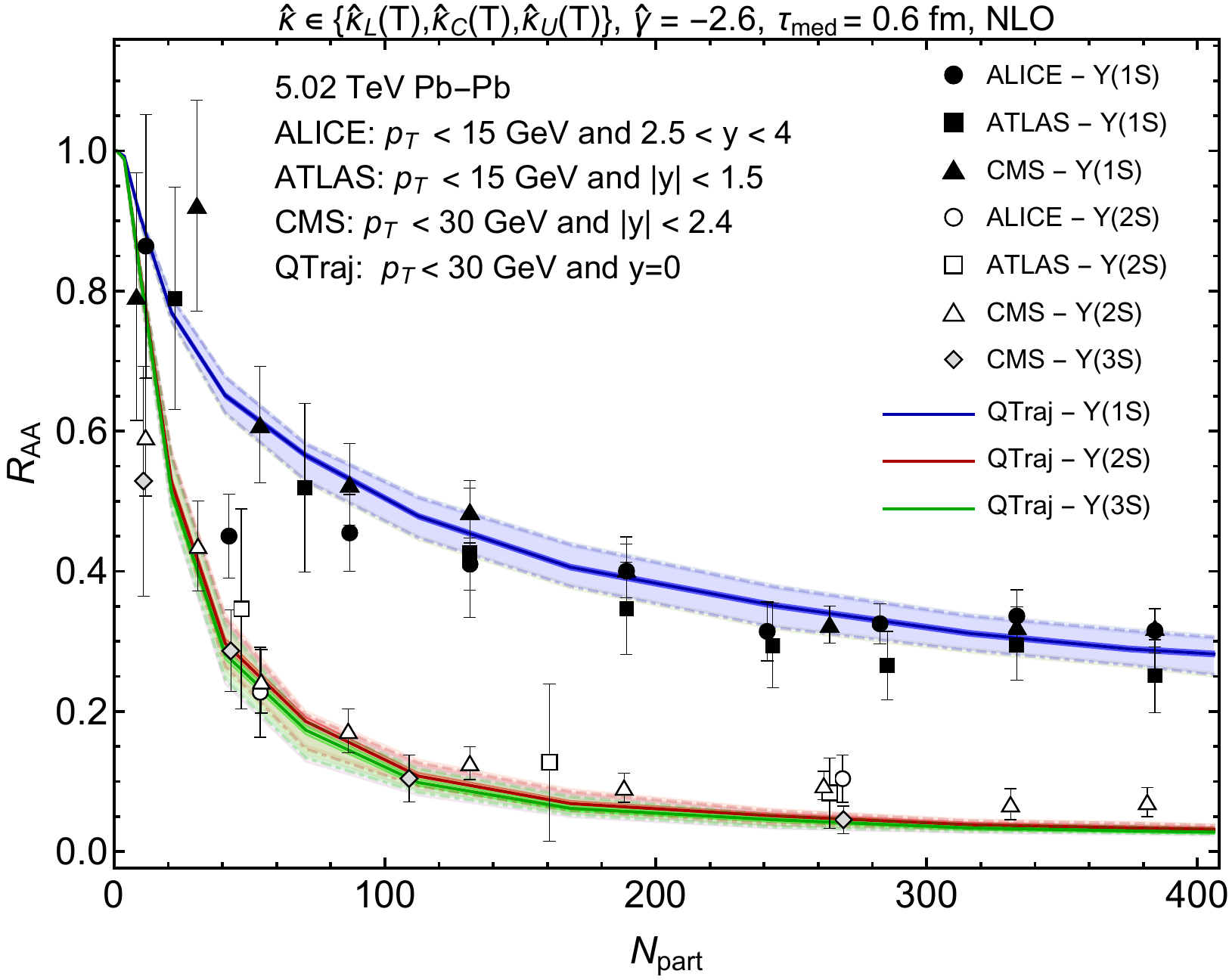} \hspace{2mm}
\includegraphics[width=0.48\linewidth]{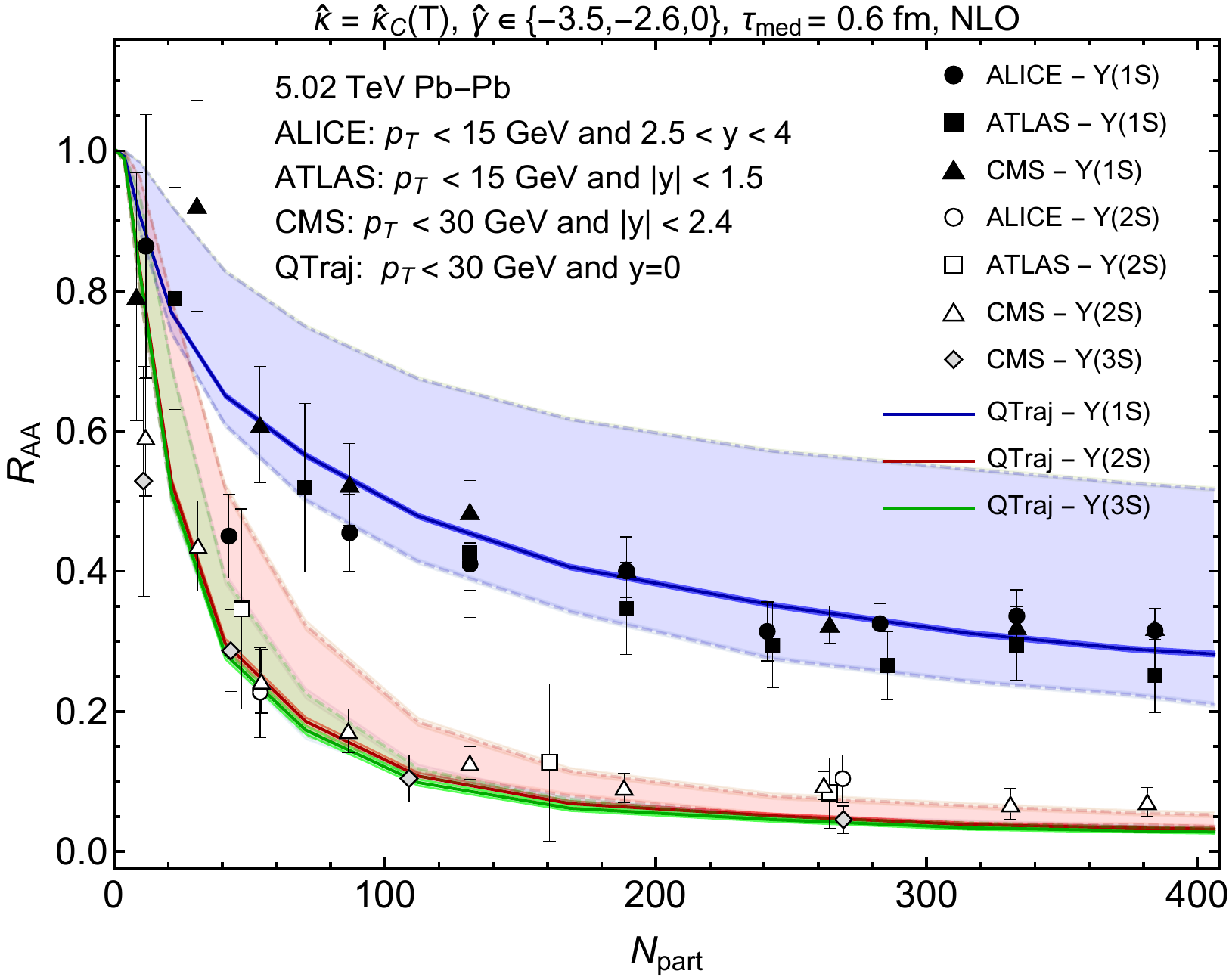}  \hspace{1mm}
\end{center}
\vspace{-5mm}
\caption{
The nuclear suppression, $R_{AA}[\Upsilon(1S,2S,3S)]$, as a function of the number of participants, $N_{\rm part}$.  The left panel shows variation of $\hat\kappa$ and the right panel shows variation of $\hat\gamma$.  The experimental results shown are from the ALICE~\cite{Acharya:2020kls}, ATLAS~\cite{Lee:2021vlb}, and CMS~\cite{Sirunyan:2018nsz,CMS:2022rna} collaborations.}
\label{fig:raavsnpart1}
\end{figure}

\begin{figure}[t]
\begin{center}
\includegraphics[width=0.48\linewidth]{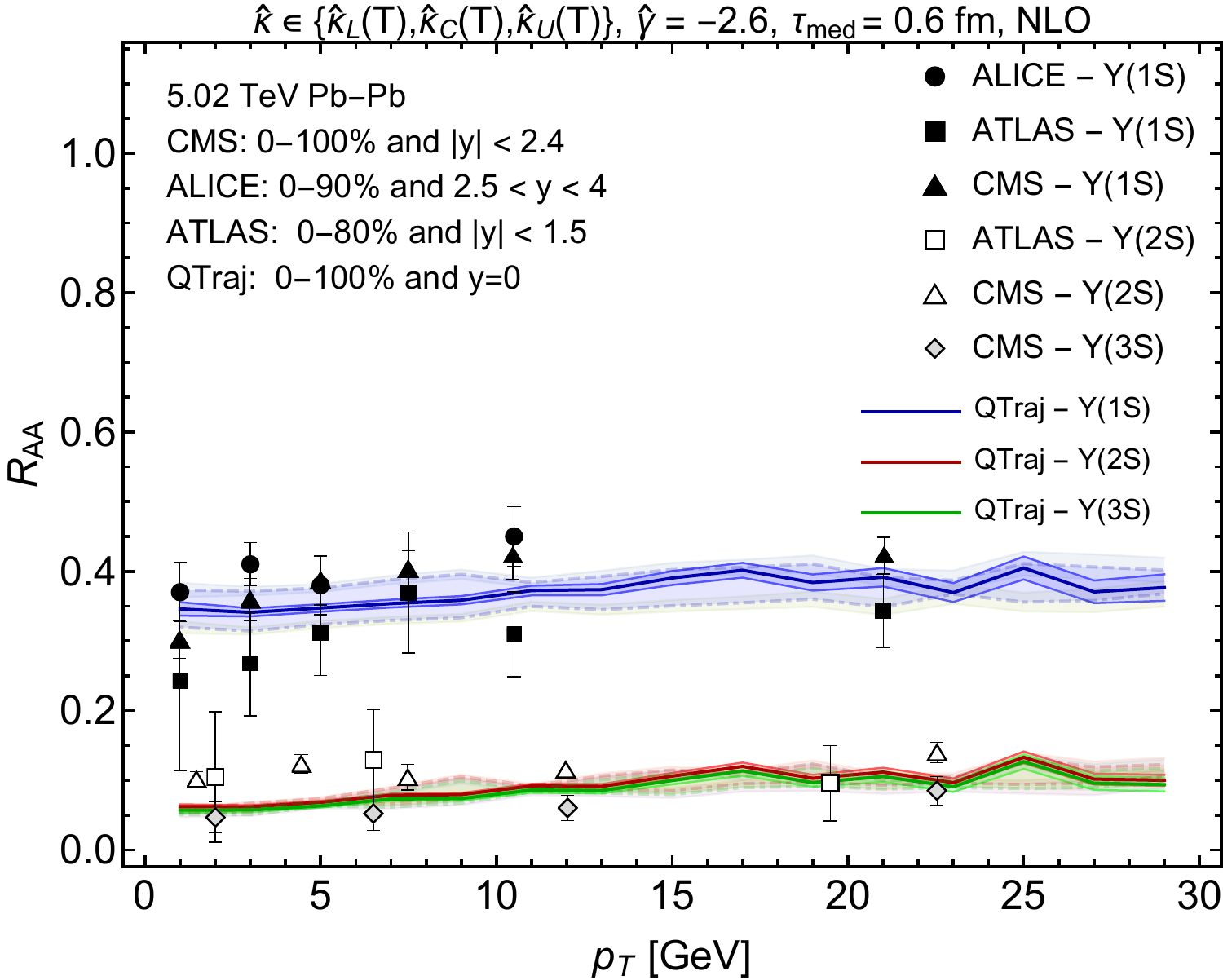} \hspace{2mm}
\includegraphics[width=0.48\linewidth]{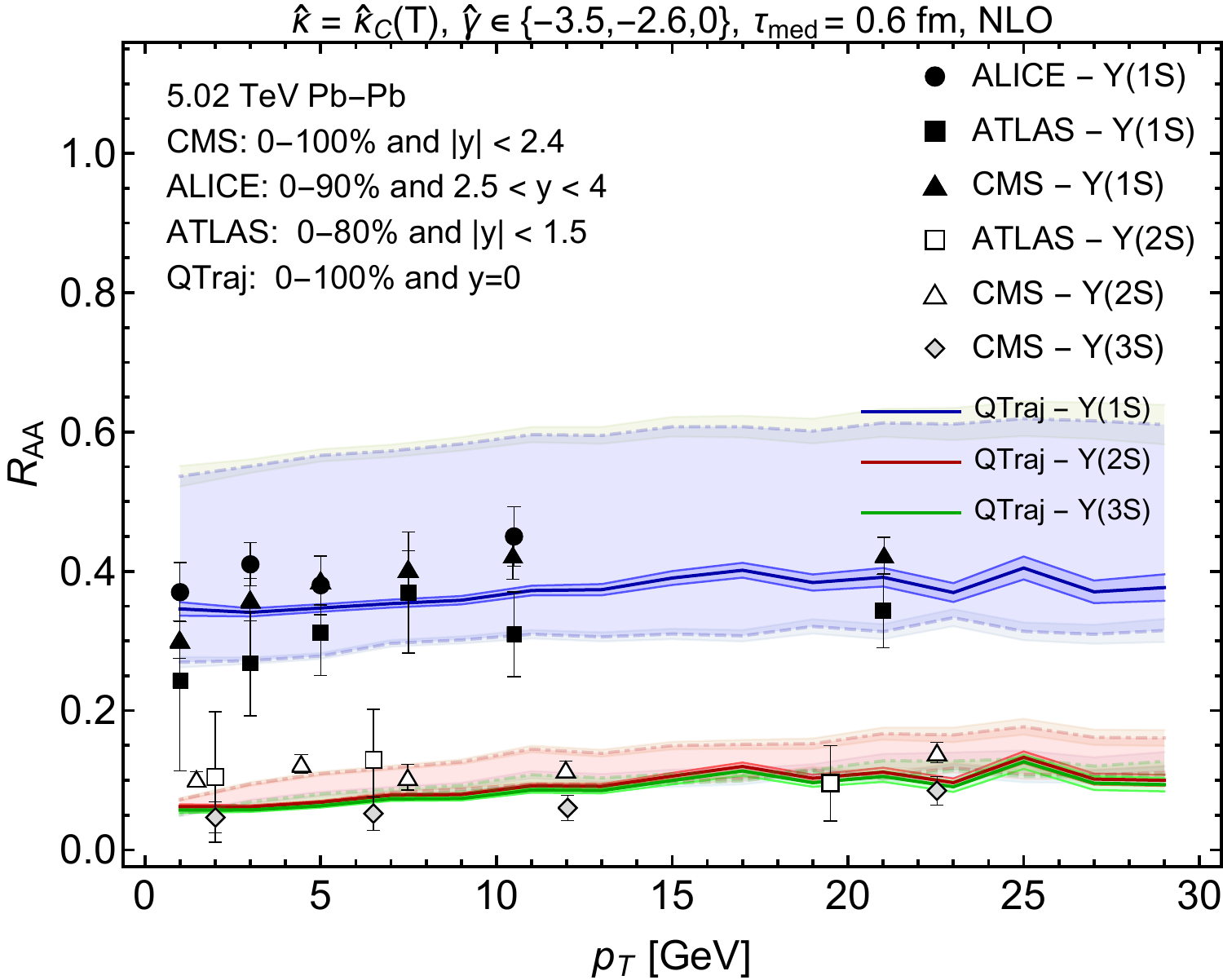} \hspace{1mm}
\end{center}
\vspace{-5mm}
\caption{
The nuclear suppression factor, $R_{AA}$, for $\Upsilon(1S,2S,3S)$ as a function of the transverse momentum, $p_T$.  The bands, etc. are the same as Fig.~\ref{fig:raavsnpart1}.
}
\label{fig:raavspt1}
\end{figure}

\begin{figure*}[t]
	\centering
	\includegraphics[width=0.48\linewidth]{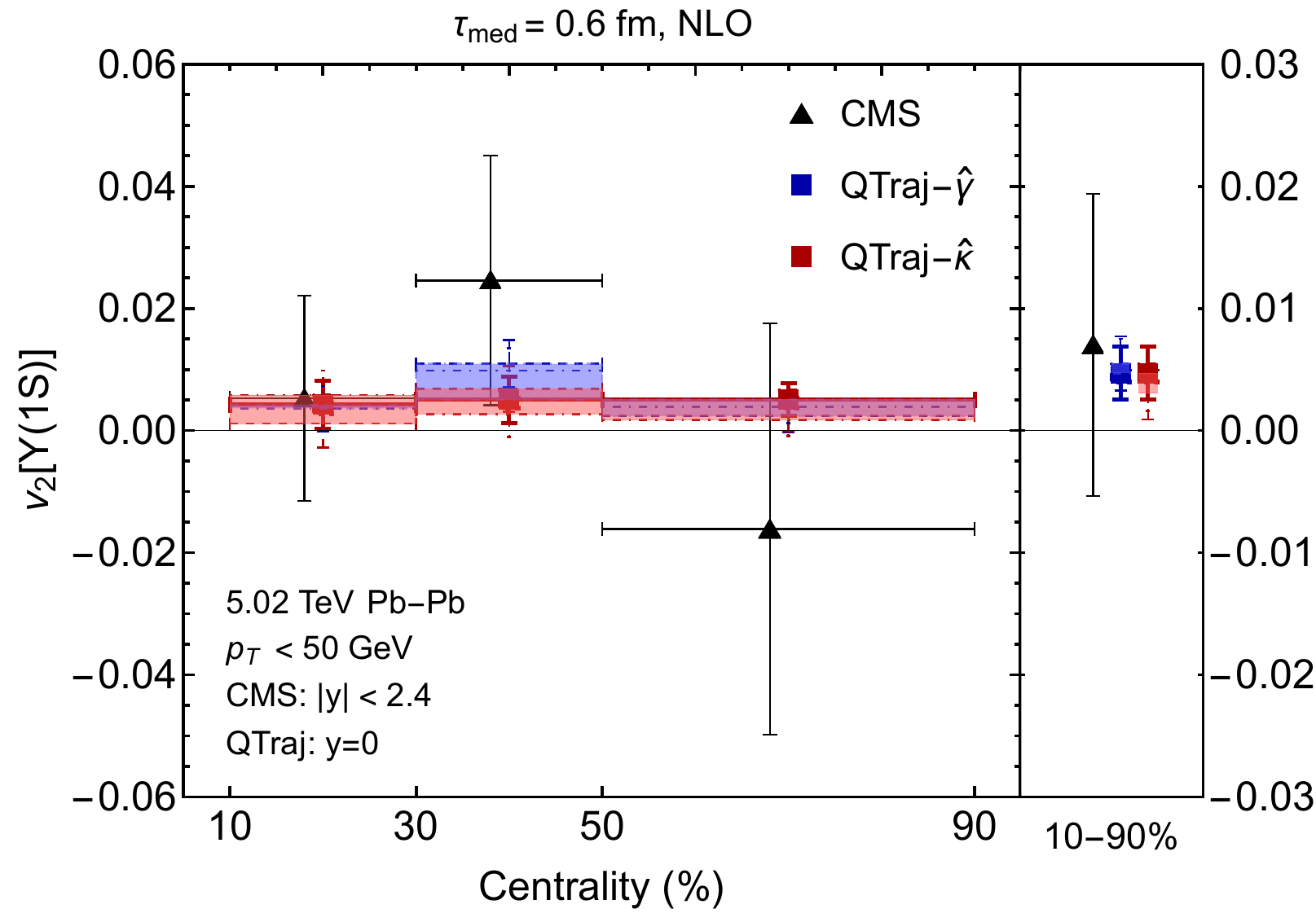} \hspace{2mm}
	\includegraphics[width=0.43\linewidth]{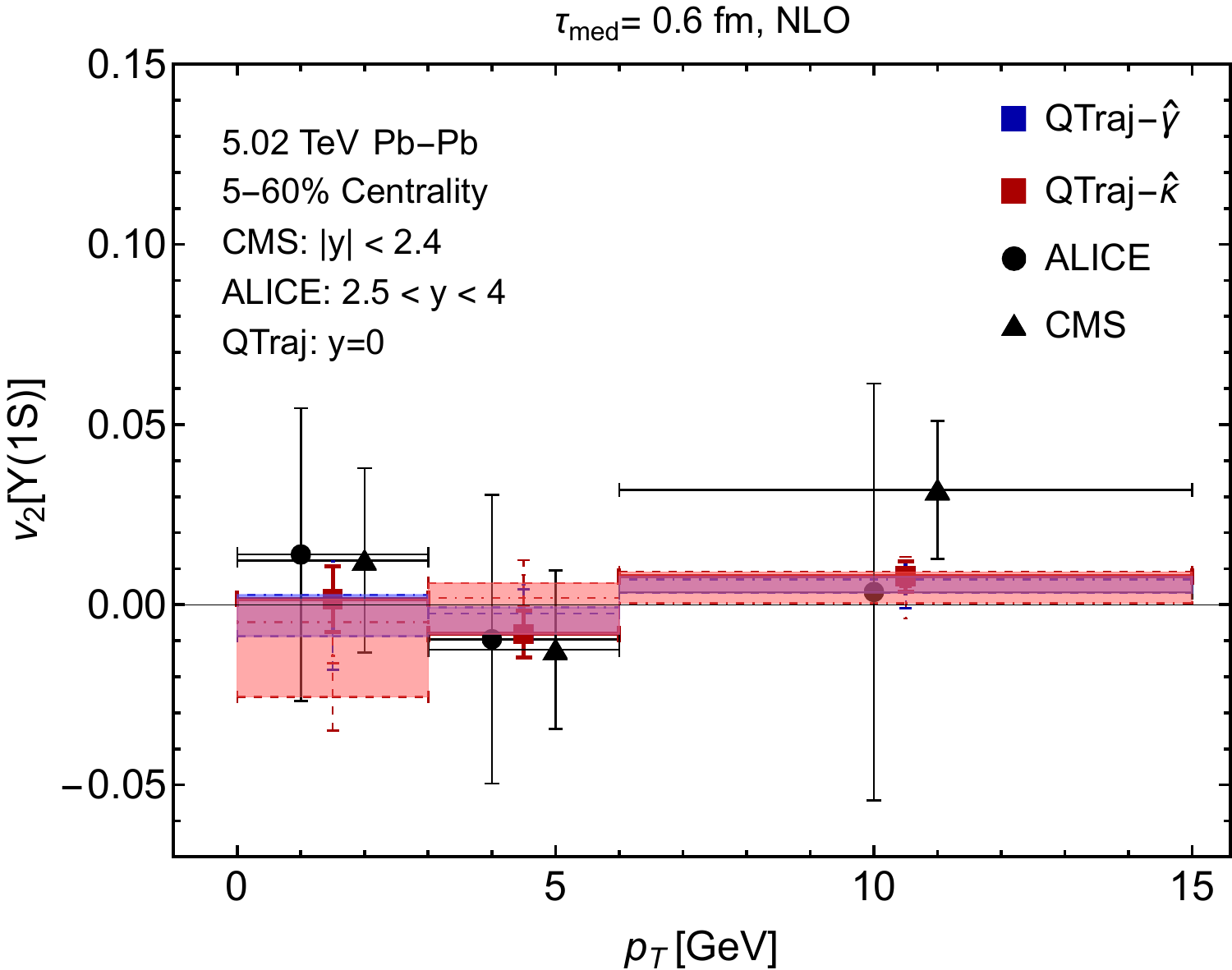} \hspace{1mm}
	\vspace{-2mm}
	\caption{The anisotropic flow coefficient $v_{2}[1S]$ as a function of centrality (left) and transverse momentum (right) obtained with fluctuating initial conditions.  We show the $\hat\gamma$ variation in blue and the $\hat\kappa$ variation in red.}
	\label{fig:v2}
\end{figure*}

For details concerning the theoretical and numerical methods employed, I refer the reader to Refs.~\cite{Brambilla:2022ynh,Alalawi:2022gul}.  The results obtained depend on two coefficients $\hat\kappa$ and $\hat\gamma$, which were extracted directly and indirectly from lattice QCD calculations \cite{Aarts:2011sm,Kim:2018yhk,Brambilla:2019tpt,Larsen:2019bwy,Brambilla:2020siz,Shi:2021qri}.  In Figs.~\ref{fig:raavsnpart1} and \ref{fig:raavspt1}, we present our NLO predictions for $R_{AA}$ as a function of $N_{\rm part}$ and $p_T$, respectively.    For these results we did not include the effect of dynamical quantum jumps.  In the left panel we show the variation of $\hat\kappa$ in the range $\hat\kappa \in \{ \hat\kappa_L(T), \hat\kappa_C(T), \hat\kappa_U(T) \}$ while holding $\hat\gamma = -2.6$.  
This value of $\hat\gamma$ was chosen as to best reproduce the $R_{AA}[\Upsilon(1S)]$.  
In the right panel we show the variation of $\hat\gamma$ in the range $-3.5 \leq \hat\gamma \leq 0$ with $\hat\kappa(T) = \hat\kappa_C(T)$.  The solid line corresponds to $\hat\gamma = -2.6$.  As this figure demonstrates, our NLO predictions without quantum jumps are in quite good agreement with the experimental data for $R_{AA}[1S]$ and $R_{AA}[3S]$.  However, for the 2S excited state, our NLO predictions without quantum jumps are somewhat lower than the experimental results, particularly for the most central collisions.

Recently, we computed the NLO bottomonium $R_{AA}$ and $v_2$ using both smooth and fluctuating initial conditions for the hydrodynamic evolution~\cite{Alalawi:2022gul}.  In Ref.~\cite{Alalawi:2022gul} in was demonstrated that the results for $R_{AA}$ obtained using fluctuating and smooth initial conditions were nearly identical, indicating that initial state fluctuations do not play an important role in this observable.    In Fig.~\ref{fig:v2}, I present the OQS+pNRQCD+IP-Glasma predictions for $v_{2}[1S]$ as a function of centrality (left panel) and transverse momentum (right panel) compared with experimental data from the ALICE and CMS collaborations \cite{ALICE:2019pox,CMS:2020efs}.  From this figure we see that the NLO OQS+pNRQCD+IP-Glasma framework predicts a rather flat dependence on centrality, with the maximum $v_2[1S]$ being on the order of 1\%.  In the right portion of the left panel, we present the results integrated over centrality as two points that include the observed variations with $\hat\kappa$ and $\hat\gamma$, respectively.\footnote{The scale of the right portion of the left panel is different from the left portion of this panel in order to make it more readable.}  The size of the error bars reflects the statistical uncertainty associated with the double average over initial conditions and physical trajectories~\cite{Alalawi:2022gul}.  The red and blue shaded regions correspond to the uncertainty associated with the variation of $\hat\kappa$ and $\hat\gamma$, respectively.  Finally, in the right panel of Fig.~\ref{fig:v2}, I present the dependence of $v_2[1S]$ on transverse momentum.

\section{Conclusions}

In this proceedings contribution, I focused on recent research that uses an OQS framework applied within the pNRQCD effective field theory.  I presented predictions for the nuclear suppression factor ($R_{AA}$) and elliptic flow coefficient ($v_2$) based on smooth and fluctuating hydrodynamical initial conditions. We found that the impact of fluctuating initial conditions was small when considering $R_{AA}$, but a larger, though still within statistical uncertainties, effect was observed for $v_2$. For $R_{AA}[1S]$, $R_{AA}[3S]$, and $v_2[1S]$, we found good agreement between the NLO OQS+pNRQCD framework and experimental data. However, we found that the amount of $\Upsilon(2S)$ suppression was slightly overestimated regardless of the hydrodynamic initial conditions used.

\setlength{\bibsep}{0.3ex}
\small
\bibliographystyle{apsrev}
\bibliography{strickland}

\end{document}